\RequirePackage{fix-cm}
\documentclass[natbib,smallextended]{svjour3}       
\smartqed  
\usepackage{graphicx}
\usepackage{color}
\usepackage{epsfig,amsmath}
\usepackage{times}
\newcommand{\chandra}{{\it Chandra}}
\newcommand{\spitzer}{{\it Spitzer}}
\newcommand{\fermi}{{\it Fermi}}
\newcommand{\xmm}{{\it XMM-Newton}}
\newcommand{\Kep}{K_\mathrm{ep}}
\newcommand{\lsim}{\raise0.3ex \hbox{$<$\kern-0.75em\raise-1.1ex\hbox{$\sim$}}\,}
\newcommand{\gsim}{\raise0.3ex \hbox{$>$\kern-0.75em\raise-1.1ex\hbox{$\sim$}}\,}
\newcommand{\gamray}{$\gamma$-ray}

\newcommand{\Emax}{$E_{\rm max}$}

 \journalname{svjour3}
\begin{document}

\title{Supernova remnants interacting with molecular clouds: X-ray and gamma-ray signatures
}

\titlerunning{SNRs interacting with molecular clouds}        

\author{Patrick Slane \and Andrei Bykov \and Donald~C.~Ellison \and Gloria Dubner \and Daniel~Castro
}


\institute{P. Slane \at
Harvard-Smithsonian Center for Astrophysics, 60 Garden St., Cambridge, MA 02138 USA \\
\email{slane@cfa.harvard.edu} 
\and
A.M. Bykov \at
A.F. Ioffe Institute for Physics and Technology, 194021, St. Petersburg, Russia \\
and\\
St. Petersburg State Polytechnical University, St. Petersburg, Russia\\
\email{byk@astro.ioffe.ru}
\and
D. C. Ellison \at
Physics Department, North Carolina State University, Box 8202, Raleigh, NC 27695, USA \\
\email{don\_ellison@ncsu.edu}
\and
G. Dubner \at
Instituto de Astronom\'{i}a y F\'{i}sica del Espacio (IAFE), UBA-CONICET, CC 67, Suc. 28, 1428 Buenos Aires, Argentina\\
\email{gdubner@iafe.uba.ar}
\and
D. Castro \at
MIT-Kavli Center for Astrophysics and Space Research, 77 Massachusetts Avenue, Cambridge, MA 02139, USA \\
\email{castro@mit.edu}
}

\date{Received: date / Accepted: date}

\maketitle

\begin{abstract}

The giant molecular clouds (MCs) found in the Milky Way and similar
galaxies play a crucial role in the evolution of these systems. The
supernova explosions that mark the death of massive stars in these
regions often lead to interactions between the supernova remnants
(SNRs) and the clouds. These interactions have a profound effect
on our understanding of SNRs. Shocks in SNRs should be capable of
accelerating particles to cosmic ray (CR) energies with efficiencies
high enough to power Galactic CRs.  X-ray and $\gamma$-ray studies
have established the presence of relativistic electrons and protons
in some SNRs and provided strong evidence for diffusive shock
acceleration as the primary acceleration mechanism, including
strongly amplified magnetic fields, temperature and ionization
effects on the shock-heated plasmas, and modifications to the
dynamical evolution of some systems. Because protons dominate the
overall energetics of the CRs, it is crucial to understand this
hadronic component even though electrons are much more efficient
radiators and it can be difficult to identify the hadronic component.
However, near MCs the densities are sufficiently high to allow the
$\gamma$-ray emission to be dominated by protons. Thus, these
interaction sites provide some of our best opportunities to constrain
the overall energetics of these particle accelerators. Here we
summarize some key properties of interactions between SNRs and MCs,
with an emphasis on recent X-ray and $\gamma$-ray studies that are
providing important constraints on our understanding of cosmic rays
in our Galaxy.

\end{abstract}


\section{Introduction}
\label{intro}

Within star-forming galaxies, cold molecular material forms complex
structures spanning a large range of spatial scales and densities,
from giant molecular clouds (MCs) spanning tens of parsecs down to
compact self-gravitating cores from which stars may form. For the
most massive stars, their short lifetimes will not provide sufficient
time to stray far from the molecular environments in which they
were born before they undergo supernova explosions. We thus expect
to find many supernova remnants (SNRs) associated with core-collapse
supernovae to evolve in regions with dense MCs.
Interactions with MCs can strongly impact the properties and long-term
evolution of SNRs, and the distinct signatures of SNR/MC interactions
can place strong constraints on this process, and on the properties
of the evolving plasma. These interactions can drive shocks into
the molecular material, providing distinct signatures of the
interaction, and can play a role in triggering new star formation.
In addition, the presence of dense gas from MC regions can play a
crucial role in revealing the presence of energetic ions accelerated
by SNR shocks, through the production of \gamray s.

The evolution of an SNR is arguably most readily investigated through
X-ray observations. The rapid shocks heat the ambient medium and
supernova (SN) ejecta to temperatures of $\sim 10^7$~K, producing thermal
bremsstrahlung and line emission from which the density, ionization state,
and abundances of the shocked gas can be constrained. Comparison of these
derived properties with models for SNR evolution produce some of the
most detailed information about the progenitor systems and their
environments. At the same time, efficient particle acceleration often
occurs in fast SNR shocks, resulting in X-ray synchrotron radiation
in the compressed and amplified magnetic fields. The relativistic
protons and electrons produced in this acceleration process can produce
\gamray\  emission. In particular, emission associated with the decay
of neutral pions produced in proton-proton collisions can be significantly
enhanced in environments with high densities, such as those in which
SNRs are interacting with MCs. Thus, X-ray and \gamray\ signatures from
SNRs carry particularly important information on interactions with MCs.

\section{Observational Signatures of SNR/MC Interactions}
\noindent
Resolved studies of SNRs and their associated clouds are most easily
carried out for systems within the Milky Way, though with the
increasing capacity of new observational facilities, the studies
are rapidly extending to our neighbor galaxies, the Magellanic
Clouds.  In our Galaxy, molecular gas clouds account for less than
one percent of the volume of the interstellar medium (ISM), yet as
they are the densest part of the medium, they represent roughly
one-half of the total gas mass interior to the Solar circle. This
mass is distributed predominantly along the spiral arms and within
a narrow midplane with a scale-height $Z \sim 50-75$~pc, much thinner
than the atomic or ionized gas \citep{cox05}. Most of the molecular
gas ($\sim$ 90\%) appears to be in massive structures distributed
in large clumps (Giant Molecular Clouds,GMCs), filaments and condensed
clumps (which are the cradle for future stars).  As mentioned above,
most core-collapse SNe are located close to molecular concentrations,
their birth places. Therefore a large percentage of the few hundreds
of observed Galactic SNRs is expected to be physically related to,
and maybe interacting with, MCs.

However, to unambiguously establish whether an SNR is physically
associated with a MC, removing confusion introduced by unrelated
gas along the line of sight is not trivial. Several distinct criteria
can be used to  demonstrate a possible SNR/MC interaction.
Morphological traces along the periphery of the SNRs such as arcs
of gas surrounding parts of the SNR or indentations in the SNR outer
border encircling dense gas concentrations can be observed in images
of SNRs (see Section 3.1.1).  Usually such features indicate that
a dense external cloud is  disturbing an otherwise spherically
symmetric shock expansion.  These initial signatures need to be
confirmed with more convincing, though more rare, features like
broadenings, wings, or asymmetries in the the molecular line spectra
\citep{fm98}, high  ratios between lines of different excitation
state  \citep{seta_etal_98}, detection of near infrared H$_2$ or
[Fe II] lines \citep[e.g.,][]{reach_etal_05}, peculiar infrared
colors \citep{reach_etal_06,Castelletti_etal_11}, or the presence
of OH (1720) MHz masers
\citep[e.g.,][]{frail_etal_96,green_etal_97,claussen_etal_97,hewitt_etal_08},
the most powerful tool to diagnose SNR/MC interactions. The fulfillment
of one or more of these different criteria can be used to propose
the existence of ``definitive", ``probable" or ``possible" physical
interactions between an SNR and a neighboring cloud.

Once the association between an SNR and a molecular feature is
firmly established, it serves to provide an independent estimate
for the distance to the SNR through the observed Doppler shift of
the line centroid and  by applying a circular rotation model for
the Galaxy.\footnote{The drawback of this method is that Galactic
circular rotation is an approximation only valid for gas close to
the Galactic disk, and for sources located inside the solar circle
there is an ambiguity of two different distances producing the
same radial velocity.} Based on a combination of different techniques,
\citet{cj13} presented a list of  $\sim 70$ Galactic SNRs
suggested to be physically interacting with neighboring MCs, of
which 34 cases are confirmed on the basis of  simultaneous fulfillment
of various criteria,  11 are probable, and 25  are classified as
possible and deserve more studies.

The basic tool used to investigate cases of SNR/MC interactions is
the survey of the interstellar medium in a field around the SNR
using different spectral lines, from the cold, atomic hydrogen
emitting at $\lambda$ 21 cm to the dense shielded regions of molecular
hydrogen (that constitutes at least 99\% of the molecular gas in
the Milky Way) emitting in the millimeter and infrared ranges.
The most widely used proxy to track down molecular gas is CO. This
molecule has a nonzero dipole moment, and radiates much more
efficiently than the abundant H$_2$ (with no dipole moment) and can
be detected easily. However, such an indirect tracer can be biased,
since there can still be CO-dark H$_2$ gas. Recent studies conducted
with the Herschel Space Observatory  revealed that the reservoir
of molecular gas in our Galaxy can be hugely underestimated when
it is traced with traditional methods \citep{pineda_etal_13}.
Through the detection of ionized carbon [CII] at 158 $\mu$m one can
trace the envelope of evolved clouds as well as clouds that are in
transition between atomic and molecular.  This might help to solve
the question about the real proximity between clouds and the CR
accelerator in some SNRs.

In what follows we discuss different tools for exploring interactions
between SNRs and MCs.

\subsection{HI Emission}

Since the discovery of the $\lambda$ 21 cm line of atomic hydrogen
more than 60 years ago, it has been shown to be the perfect tool
for investigating the distribution and kinematics of the interstellar
medium, revealing Galactic arms, large concentrations, arcs, and
bubbles. It is, then, the basic observational resource to explore
the environs of SNRs searching for candidate structures with which
the SNRs may be interacting. However, because of the high abundance
and ubiquity, confusion with unrelated gas along the line of sight
dominates, and the detection of an HI candidate structure needs to
be confirmed with other indicators that reinforce the hypothesis
of association. The HI can be studied either in emission or in
absorption, the latter being a very effective tool for constraining
the distance to the SNR.  In addition, the HI mapping of large
fields around SNRs is an excellent tool to explore the history of
the precursor star, for example by detecting large wind-blown bubbles
around the SNRs, and to estimate the gas density of the medium where
the blast wave expands.

Numerous HI studies around SNRs have been carried out using single-dish
and interferometric radio telescopes. Some examples are the results
presented on the Lupus Loop \citep{cd82}, G296.5+10.0
\citep{dubner_etal_86}, Vela \citep{dubner_etal_98a}, W50 - SS 443
\citep{dubner_etal_98b}, Tycho \citep{reynoso_etal_99}, SN~1006
\citep{dubner_etal_02}, W28 \citep{velazquez_etal_02}, Puppis~A
\citep{reynoso_etal_03}, Kes~75 \citep{lt08}, IC~443 \citep{lee_etal_08},
and several Southern Galactic SNRs \citep{koo_etal_04}. Recent
observations also indicate the existence of shells and super-shells
around SNRs \citep{park_etal_13}, which are particularly useful
as they permit the reconstruction of the history of the SNR progenitor,
providing hints on the nature of of ``dark" \gamray\ sources
\citep[e.g.,][]{gabanyi_etal_13}.

\subsection{Molecular Emission}

As mentioned above CO studies are the most widely used tool to
analyze distribution and kinematics of cold, dense clouds with high
molecular content. $^{12}$CO in its different excitation states and
$^{13}$CO lines have been surveyed over most of the sky and public
data are available from the classical survey ``The Milky Way in
Molecular Clouds'' by
\citet{dame_etal_01}\footnote{http://www.cfa.harvard.edu/mmw/MilkyWayinMolClouds.html},
the ``FCRAO CO Survey of the Outer Galaxy'' by
\citet{heyer_etal_98}\footnote{http://www.astro.umass.edu/~fcrao/telescope/2quad.html},
the ``Galactic Ring Survey'' by
\citet{jackson_etal_06}\footnote{http://www.bu.edu/galacticring/}, or
the new ``MOPRA Southern Galactic Plane CO Survey'' by
\citet{burton_etal_13}, among others. These public databases are
an excellent starting point to search for possible SNR/MC interactions.
Additionally, dedicated studies using different facilities and in
different molecular transitions have been conducted towards many Galactic
SNRs.  The SNR IC443 is a textbook case to analyze shock chemistry,
from the early work by \cite{DF81}, to tens of works investigating the
chemical and physical transformations introduced by the strong SNR
shocks on the surrounding MCs
\citep[e.g.,][]{white_etal_87,burton_etal_88,ws92,vanD_etal_93}.
More recently, Castelletti et al. (2011) showed a comparison
between very high energy \gamray\ emission as detected with VERITAS
\citep{acciari_etal_09}, with $^{12}$C0 J=1-0 integrated emission
\citep{zhang_etal_10} revealing for the first time the excellent
concordance between emissions in both regimes.

\begin{figure}[t]
\includegraphics[height=0.4\textheight]{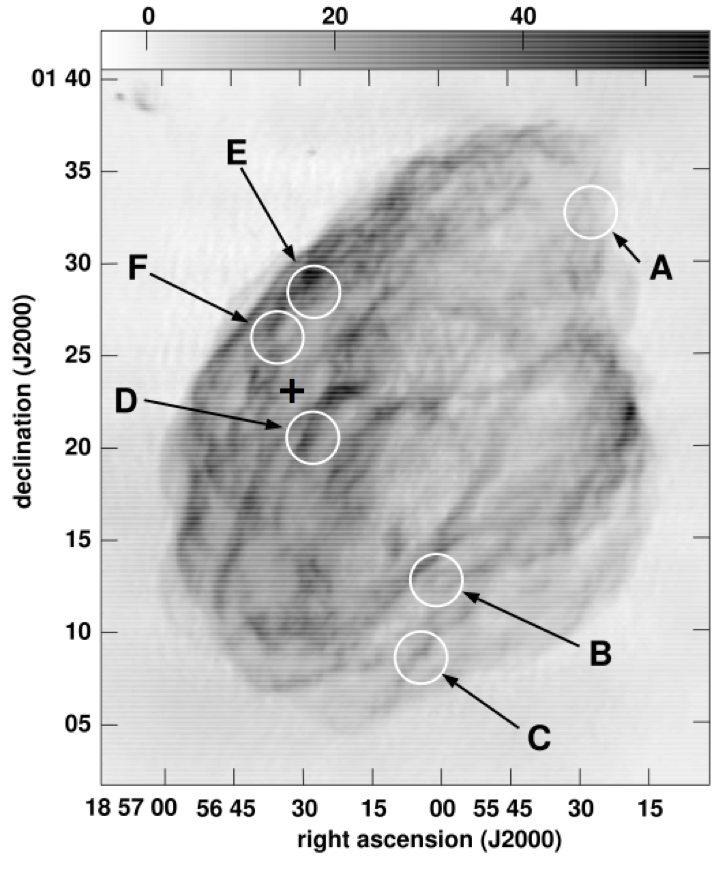}
\includegraphics[height=0.4\textheight]{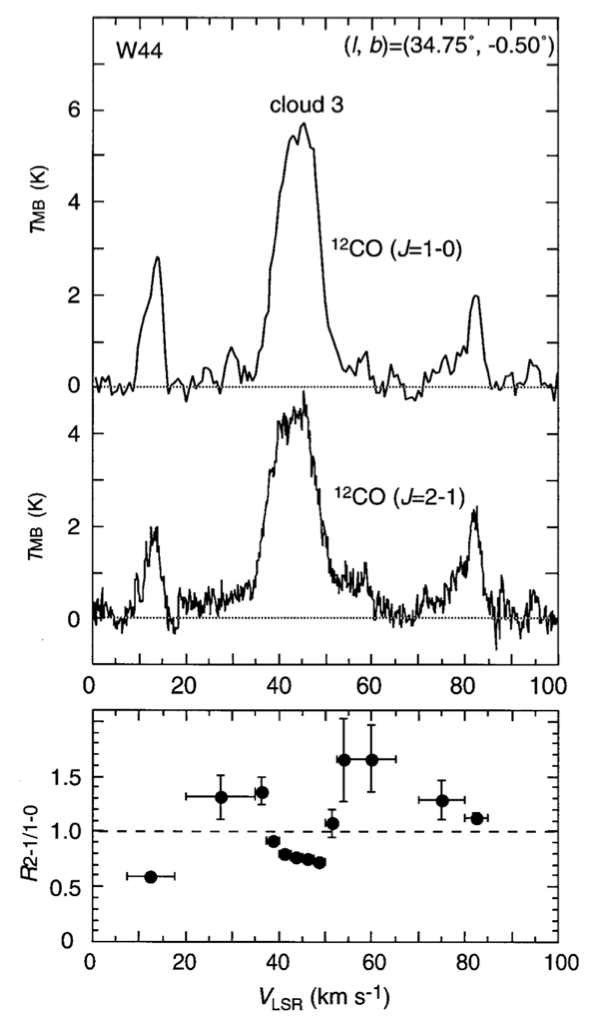}
\caption{
Left: VLA image of W44. Circles indicate the positions of observed
OH maser emission, and cross indicates the position from which the
CO profile at the right is taken.  Right: CO line profile from Cloud
C in field of W44, illustrating broad wings produced by an interaction
between the cloud and the SNR. [Left panel from Hoffman et al. 2005.
Right panel from Seta et al. 1998. Reproduced by permission of the
AAS.]
}
\label{w44}
\end{figure}

An important study carried out by \citet{seta_etal_98} proposed
that an enhanced $^{12}$CO(J=2-1)$/^{12}$CO(J=1-0) ratio in the
line wings is a clear signature of physical interaction. Ratios
are $\geq 3$ in IC443, and higher than $\sim 1$ in W44, which also
shows broadened line profiles indicating disrupted molecular 
material (Figure \ref{w44}, right).

\citet{ht86} carried out a search for possible SNR/MC associations
in the outer Galaxy, within a limited region of the northern
hemisphere, finding about 13  possible associations based on spatial
coincidences (out of 26 SNRs investigated).  More recently,
\citet{chen_etal_13} reported CO observations towards a number of
SNRs possibly interacting with MCs.

\subsection{Masers}

OH (1720 MHz) masers have long been recognized as signposts for SNR/MC
interactions \citep[e.g.,][]{frail_etal_94}. This maser line is
inverted through collisions with H$_2$ behind non-dissociative
C-type shocks propagating into MCs \citep{fm98,rr99}.
The conditions required for the maser formation are rather narrow,
with temperatures in the range $50 - 125$~K, densities between $n
= 10^3 - 10^5 {\rm\ cm}^{-3}$, and OH column densities of $10^{16}
- 10^{17} {\rm\ cm}^{-2}$ \citep{lockett_etal_99}. The large column
requirement exceeds what can be produced by the slow shocks,
indicating that an additional contribution is required to dissociate
water into OH \citep{wardle99}, possibly from X-rays from the hot
SNR shell or from cosmic rays that may have been accelerated by the
SNR. As the physical conditions needed to pump this maser are so
strict, it has to be noted that their presence is sufficient to
demonstrate interaction, but its absence does not rule it out.

To date, OH (1720 MHz) masers have been found in $\sim$ 24 SNRs,
or 10\% of the known SNRs in our Galaxy 
\citep{brogan_etal_13}.
As summarized by \citet{brogan_etal_13}, SNR OH (1720
MHz) masers have a large maser spot size, narrow and simple line
profiles, low levels of circular polarization (less than 10\%), and
low magnetic field strength.

The location of masers show, in general, very good coincidence with
density/shock tracers. In the case of W44
\citep{frail_etal_96,claussen_etal_97,hoffman_etal_05}, there is a
strong correlation between the morphology of the molecular gas and
the relativistic gas traced by synchrotron emission at centimeter
wavelengths.

In addition to providing a clear indication of SNR/MC interactions,
OH(1720 MHz) masers also permit an independent estimate for the
kinematic distances to the clouds, and thus for the remnants. Also,
and very important, they provide the only means of directly observing
the magnetic field strength using the Zeeman effect 
\citep{brogan_etal_00,brogan_etal_13}.

\subsection{Radiative shocks} 
The evolution of supernova shells colliding with MCs
differs from SNRs expanding in a pre-supernova wind or of  those
expanding in a homogeneous uniform ISM \citep[][]{chevalier99}.
Observations have revealed that MCs are complex structures with a
hierarchical structure of dense clumps embedded in an inter-clump
gas of moderate density $\sim$ 10 H atoms cm$^{-3}$.  The volume
filling factor of the dense clumps is typically a few percent and
the clump mass ranges from a fraction of $M_\circ$ to thousands of
$M_\circ$.  The total mass in the clumps can be comparable or more
than the mass of the inter-clump gas.

The dense MC environment results in an early end of the adiabatic
phase of the SNR expansion.  An SNR leaves the adiabatic phase and
enters the {\sl pressure driven snowplow} (PDS) stage which eventually
results in the formation of a radiative shell at an age estimated
by  \citet{cioffiea88} to be $\tau_{\rm sf} = 3.6 \times 10^{4}
E_{51}^{3/14} n_{10}^{-4/7} \zeta_m^{-5/14}$ years, where $E_{51}$
is the kinetic energy of supernova shell in units of $10^{51}$ erg,
$n_{10}$ is the ambient number density in units of 10 cm$^{-3}$,
and $\zeta_m$ is a metallicity factor of order 1 for solar abundances.
Note that the transition of the SNR to the PDS
stage is expected to occur somewhat earlier than $\tau_{\rm
sf}$.  The shock radius and velocity at the beginning of the PDS
stage are $R_{\rm pds} = 5.2 E_{51}^{2/7} n_{10}^{-3/7}
\zeta_m^{-1/7}$ pc and $v_{\rm pds}=573 E_{51}^{1/14}
n_{10}^{1/7}\zeta_m^{3/14}$ km s$^{-1}$ correspondingly.
\citet[][]{chevalier99} noted that a power law expansion in time
is a reasonable approximation for the radiative shell over a range
(5-50)$\tau_{\rm pds}$.

Radiative shocks in evolved SNRs are characterized by the efficient
cooling of the post-shock plasma by line radiation from the dense
shell behind the collisionless shock transition. The structure  of
the radiative shock depends on the line opacity which, for radiative
SNR shocks, typically allows the escape of some optical and
fine-structure infrared (IR) lines of abundant ions providing
observational diagnostics of the shocks \cite[see
e.g.,][]{raymond79,hm89,gs09,bmrkv13}.

\begin{figure}[t]
\includegraphics[height=0.265\textheight]{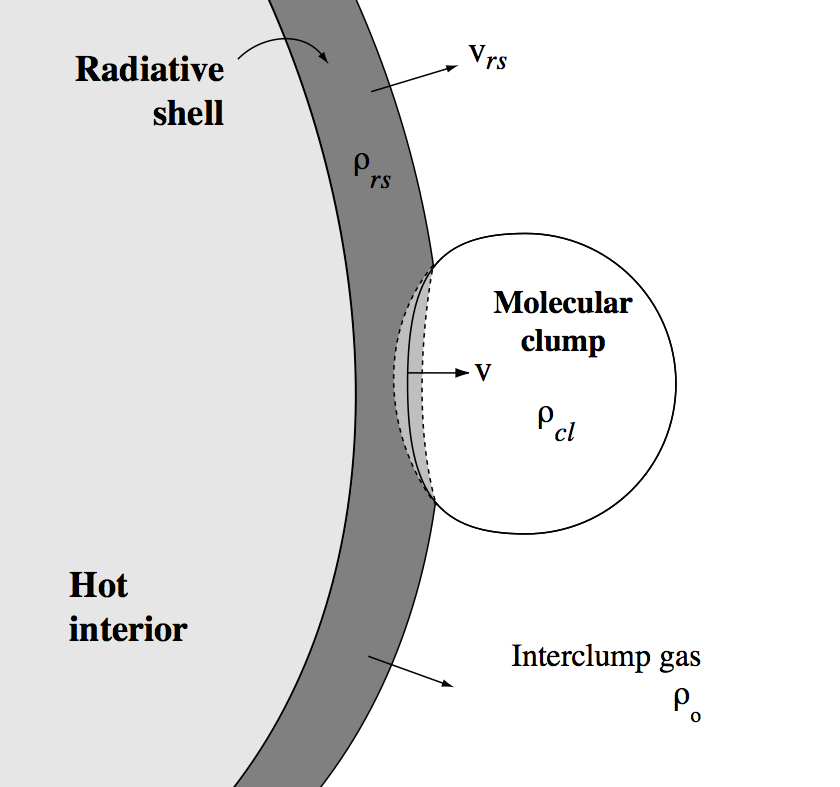}
\includegraphics[height=0.265\textheight]{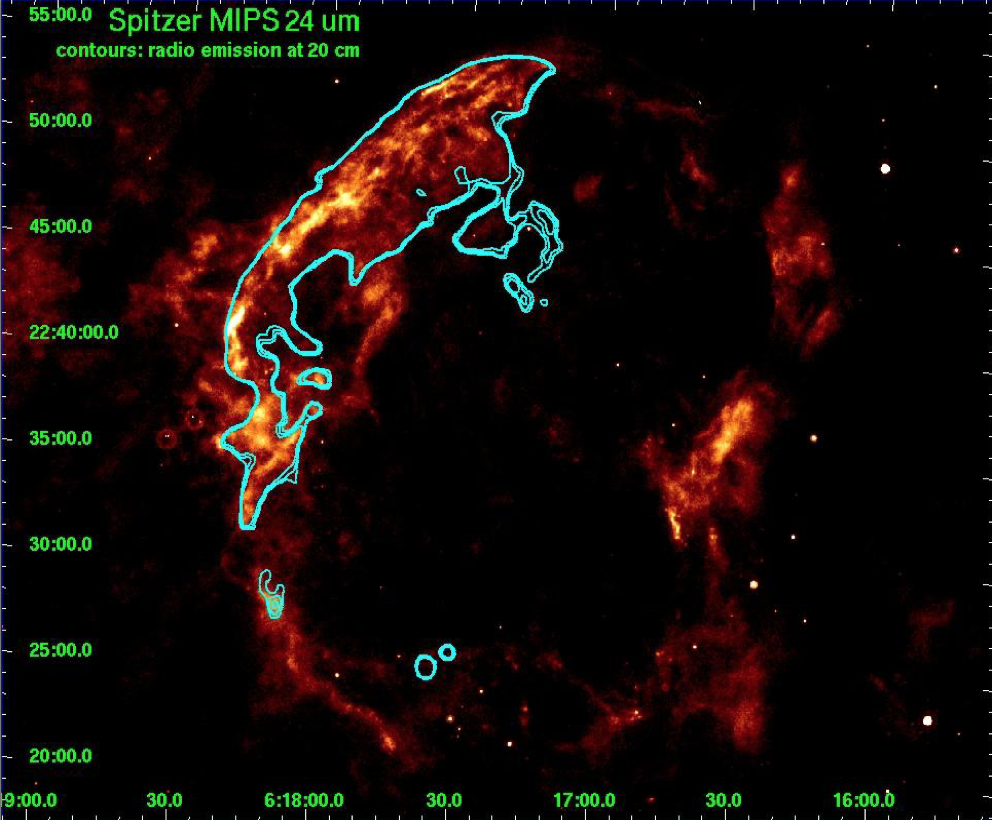}
\caption{ Left: The sketch illustrating the interaction of the
radiative supernova shell with a clumpy MC as described
by Chevalier (1999). Right: {\it Spitzer MIPS} 24 $\mu$m image of
IC~443 SNR with 1.4 GHz radio contours from Bykov et al (2008). The
radiative shock is located in the northeast part of the remnant. }
\label{rad_shell}
\end{figure}

In the case of core-collapse supernovae in such MC environments,
the remnant becomes radiative at a radius of $\lsim 6$ pc, forming
a shell that is magnetically supported and whose structure can be
described by a self-similar solution \citep[][]{chevalier99}.  The
expected structure of the radiative shell is illustrated in the
left panel in Fig.~\ref{rad_shell}. The interaction of the radiative
shell with an ensemble of molecular clumps of different sizes,
masses, and cloud magnetization in the model results in both J and
C-type shocks with a range of velocities. In some cases, OH maser
emission may be associated with shocked molecular clumps. For IC~443,
for example, \citet{snellea05} detected the shocked clumps B, C,
and G in H$_2$O, $^{13}$CO, and C I line emission with  {\sl
Submillimeter Wave Astronomy Satellite (SWAS)} detectors. They
concluded that, to explain these observations, different type shocks
with a range of velocities is likely required.  Recently,
\citet[][]{mol_clumpIC443_lee12} found observational evidence for
shocked clumps  by mapping the southern part of IC~443 with $^{12}$CO
J = 1-0 and HCO+ J = 1-0 lines.

Radiative shocks are observed in the northeast part of the SNR
IC~443 where it is very likely interacting with an HI cloud.
Spectrophotometry of  [O III], [N II], and [S II]  optical line
emission performed by \citet{fk80} is found to be consistent with
a radiative shock propagating in an inhomogeneous ISM.  Electron
densities up to 500 cm$^{-3}$ were derived from the [S II] lines
measurements. \citet{IC443_IRSF_13} presented the  near-IR
[Fe II] 1.257 $\mu$m and [Fe II] 1.644 $\mu$m line maps (30' $\times$
35') of IC 443 made with {\sl IRSF/SIRIUS}.  They found  that [Fe
II] filamentary structures exist all over the remnant, not only in
the ionic shock NE-shell, but also in a molecular shock shell and
a central region inside the shells.

Earlier, {\sl Two Micron All Sky Survey (2MASS)} images of the
entire IC~443 remnant in near-IR J (1.25 $\mu$m), H (1.65
$\mu$m), and Ks (2.17 $\mu$m) were analyzed by \citet{2mass_IC443_rho01}
who revealed some clear morphological differences between the
northeastern and southern parts.  The J- and H-band {\sl 2MASS}
emission from the NE rim was attributed mostly to different ionic
fine-structure lines, where the H-band is often dominated by [Fe
II] 1.644 $\mu$m line emission.  The {\sl 2MASS} emission in the
K$_s$-band with 2.12 $\mu$m  H$_2$ molecular line emission, indicated
a shocked molecular ridge spanning the southern region  due to
interaction of the remnant with the adjacent MC.

In the right panel of Fig.~\ref{rad_shell} we show the {\sl Spitzer
MIPS} 24 $\mu$m image of IC~443 SNR taken from \citet{ic443_spitzer08}
with 1.4 GHz radio contours. The 24 $\mu$m emission is likely
dominated by [Fe II] 26 $\mu$m line emission rather than by the
heated dust.  This was first established by \citet{iso_IC443_99}
who showed, using {\sl ISO-SWS} spectroscopy, that most of the 12
$\mu$m and 25 $\mu$m band {\sl IRAS} flux is accounted for by the
line emission from [NeII] and [FeII] rather than dust emission.
Apart from the [Fe II] 26 $\mu$m line, the other fine-structure
far-IR lines [OI] 63.2 $\mu$m and [C II] 157.7 $\mu$m are
predicted to be bright in the radiative shock models of  \citet{raymond79}
and \citet{hm89}. Indeed, Reach \& Rho (1996) estimated the luminosity
of W44 in [OI] 63 $\mu$m line emission to be about $4\times 10^{36}$
erg s$^{-1}$, which \citet{chevalier99} argued was consistent with
the radiative shock models.

For the densities inferred for IC~443, the radiative shell is
expected to form when the forward shock velocity drops below  $v_{\rm
pds} \sim$ 600 km s$^{-1}$.  For shock speeds $>v_{\rm pds}$,
efficient diffusive shock acceleration (DSA) can occur and a
substantial fraction of the shock ram pressure can be transferred
into relativistic particles and turbulent magnetic fields which may
be amplified to values well above those expected from adiabatic
compression alone.  The filamentary line emission region in the NE
part of IC~443 is clearly correlated with the bright 1.4 GHz radio
emission shell.  This is an indication of the particle acceleration
process in the shock.  In contrast to the value of $v_{\rm pds}
\sim$ 600 km s$^{-1}$, the radiative shock velocity in the NE shell
of IC~443 was estimated to be about 100 km s$^{-1}$, while W44 is
likely expanding at 150 km s$^{-1}$.  It must be noted however,
that these estimates were based on line emissivity derived in
radiative shock models which did not account for efficient CR
acceleration in the shocks.  Particle acceleration, and the escape
of high-energy CRs, result in higher compression and lower post-shock
temperature at a given shock velocity than predicted ignoring DSA.
The effect of CR acceleration on the line luminosity and ratios was
discussed by \citet{bmrkv13} who found a rather strong sensitivity
of the line ratios to the extra shock compression effect. This
implies the need for self-consistent models of radiative shocks
including the efficient production of a  non-thermal components.

\section{X-rays from SNR/MC Interactions}

\noindent As a young SNR evolves, it compresses and heats the
surrounding medium to X-ray emitting temperatures.  Where the medium
is extremely dense, however, the shock velocity is slow and the
temperature of the shocked plasma can be considerably lower. In
addition, the expansion of the remnant can be significantly impeded,
leading to a distorted morphology. The high density can lead to
rapid ionization, and can also drive a strong reverse shock into
the ejecta. The X-ray characteristics thus provide important
signatures of SNR/MC interactions.

\subsection{Morphological Effects}

In the Sedov phase of evolution, the radius of an SNR at a given
age scales as $R_{\rm SNR} \propto n_0^{-1/5}.$ Despite this weak
dependence, variations in density of a factor of five will produce
significant changes in the remnant size of about 40\% and can easily
cause noticeable modifications to the remnant shape.  Molecular
cloud environments are characterized by significant variations in
density, and the clouds themselves have densities nearly $10^2 -
10^3$ times higher than that of the typical ISM. SNR evolution in
such environments can thus lead to deviations from the spherical
morphology expected under ideal evolution in a uniform medium.

In addition, the {\it apparent} morphologies of SNRs can depend
upon the density and distribution of foreground material, through
energy-dependent absorption. The transmission of the ISM to X-rays
for example, is $e^{-\sigma N_H}$ where $\sigma$ is the photoelectric
absorption coefficient and $N_H$ is the column density of gas.
Hydrogen itself does not produce significant absorption at X-rays
energies. Rather, $N_H$ acts as a tracer of other atoms with K-shell
and L-shell transition energies in the soft X-ray band ($\sim 0.1
- 10$~keV), given some relative abundance distribution relative to
H. This has two significant effects on X-ray images of SNRs. First,
for large values of $N_H$ associated with large distances through
the ISM, the observed X-ray emission will be faint. Second, the
presence of MCs along the line of sight to an SNR will produce
excess absorption.  This can result in complete absorption from
particular regions of the SNR, changing its apparent morphology.
Because the absorption coefficient increases rapidly at low energies
\citep[e.g.,][]{mm83}, comparison of the morphology in low and high
energy bands can reveal evidence for foreground MCs. Equivalently,
spectral analysis of different regions of the SNR can reveal excess
absorption of soft photons in the regions along the line of sight
to MCs.

\begin{figure}[t]
\includegraphics[height=0.33\textheight]{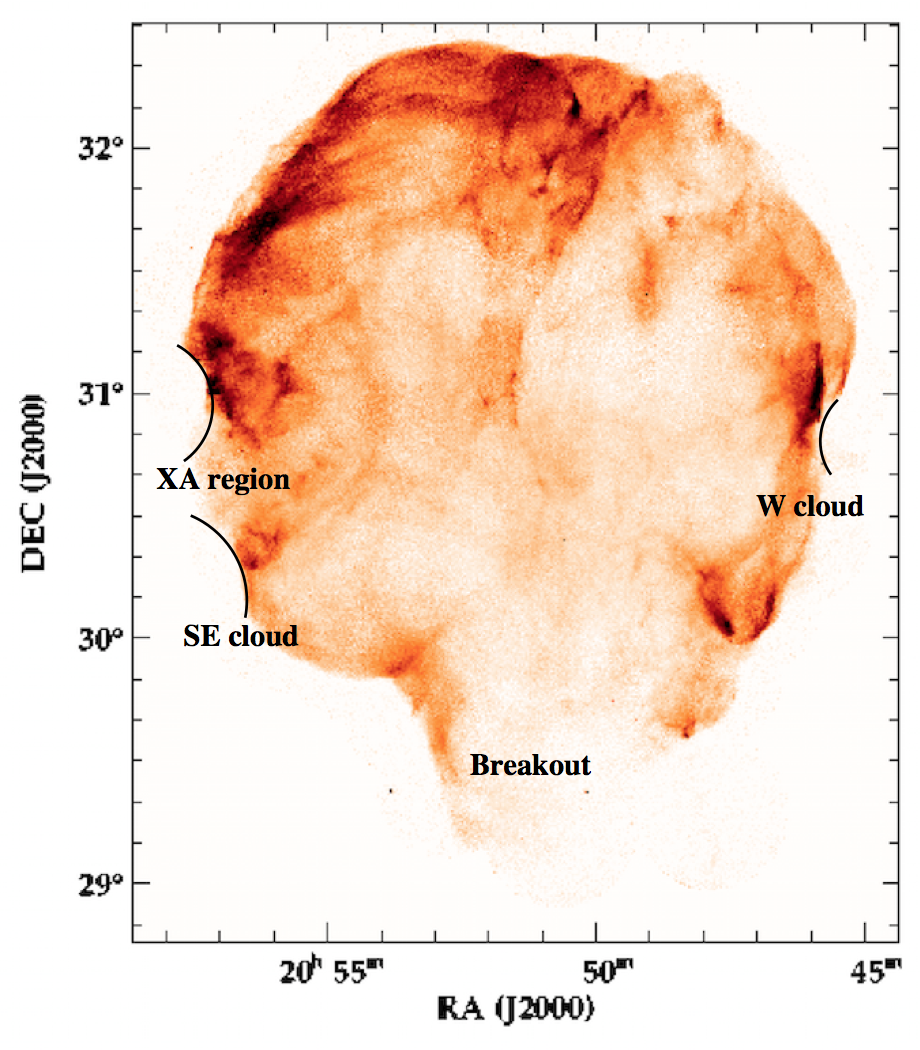}
\includegraphics[height=0.33\textheight]{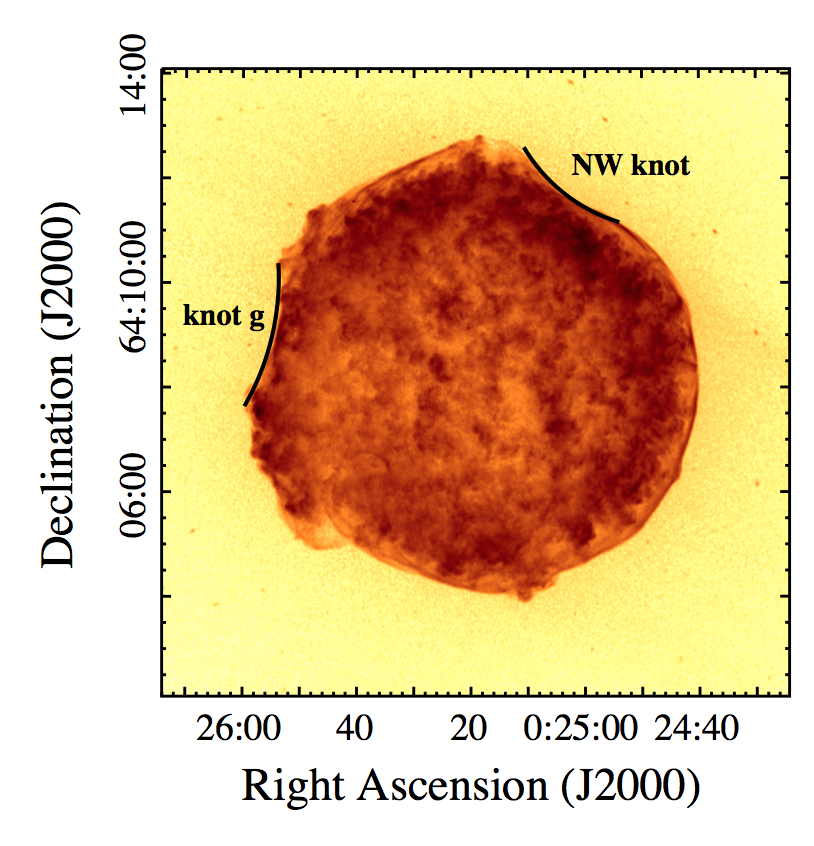}
\caption{
Left: ROSAT HRI image of Cygnus Loop. Distinct deformation of
the shell is seen at positions of cloud interactions indicated
by arcs.
Right: {\it Chandra} image of Tycho's SNR. Arcs in the east (so-called
``knot g'') and northwest indicate positions of cloud interactions.
}
\label{cloud1}
\end{figure}

\subsubsection{Shell Deformation}

The complex CSM/ISM structure surrounding many SNRs, often characterized
by significant large-scale density gradients, can result in very
significant deviations from a circular morphology
\citep[e.g.,][]{lopez_etal_09}.  Direct interactions with dense
clouds can also modify the morphology of SNRs, causing distinct
deformation of the shell-like structure. In Figure \ref{cloud1}
(left) we show the ROSAT HRI image of the Cygnus Loop
\citep{levenson_etal_99}, a nearby middle-aged SNR. The SNR shell
is roughly circular, but significant deviations from circularity
are seen in four regions. In the south, there is an apparent breakout
associated with evolution into a low-density region. In the west,
and in two positions in the east, there are very distinct arc-like
deformations known as the W knot, the SE knot, and knot XA. The SE
knot represents an encounter of the SNR blast wave with a protrusion
from a large cloud.  X-ray emission interior to radiative filaments
in this region appears to originate from the reverse shock
produced in the interaction \citep{graham_etal_95}. Knot XA appears
to correspond to a dense, clumpy region resulting from an interaction
with the wall of a cavity swept out by a precursor wind \citep{mb11}.

Also shown in Figure \ref{cloud1} is a \chandra\ image of Tycho's
SNR (right). While the overall morphology of this young, ejecta-dominated,
historical remnant (SN 1572) is fairly circular, there are obvious
depressions in the northwestern and eastern regions (indicated with
arcs in the Figure). Density estimates based on the IR emission
from the remnant shell reveal dramatic increases in these regions
\citep{williams_etal_13},
and X-ray proper motion measurements show
that the expansion rate is lower in these regions than for adjacent
regions of the remnant \citep{katsuda_etal_10}. These observations
are consistent with the interpretation that Tycho is encountering
dense clouds in the east and northwest, and optical studies of knot
g, located in the eastern limb depression, show direct evidence of
radiative shocks from the interaction.

\begin{figure}[t]
\includegraphics[height=0.37\textheight]{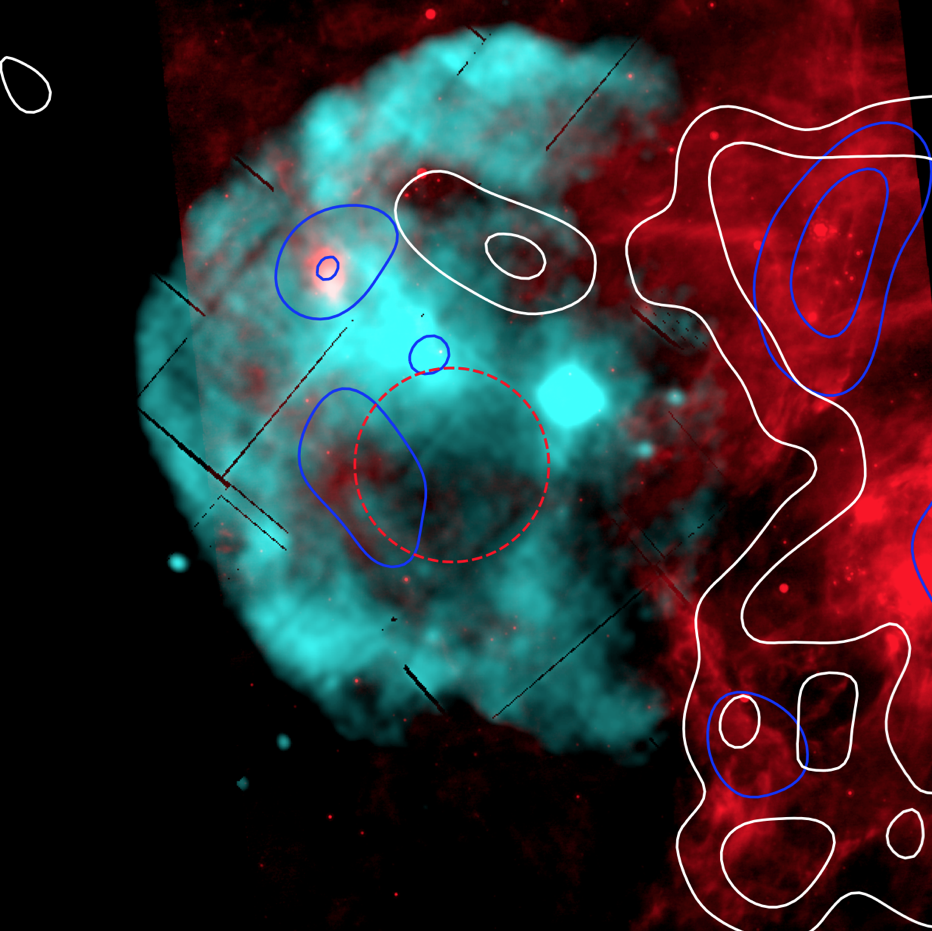}
\includegraphics[height=0.37\textheight]{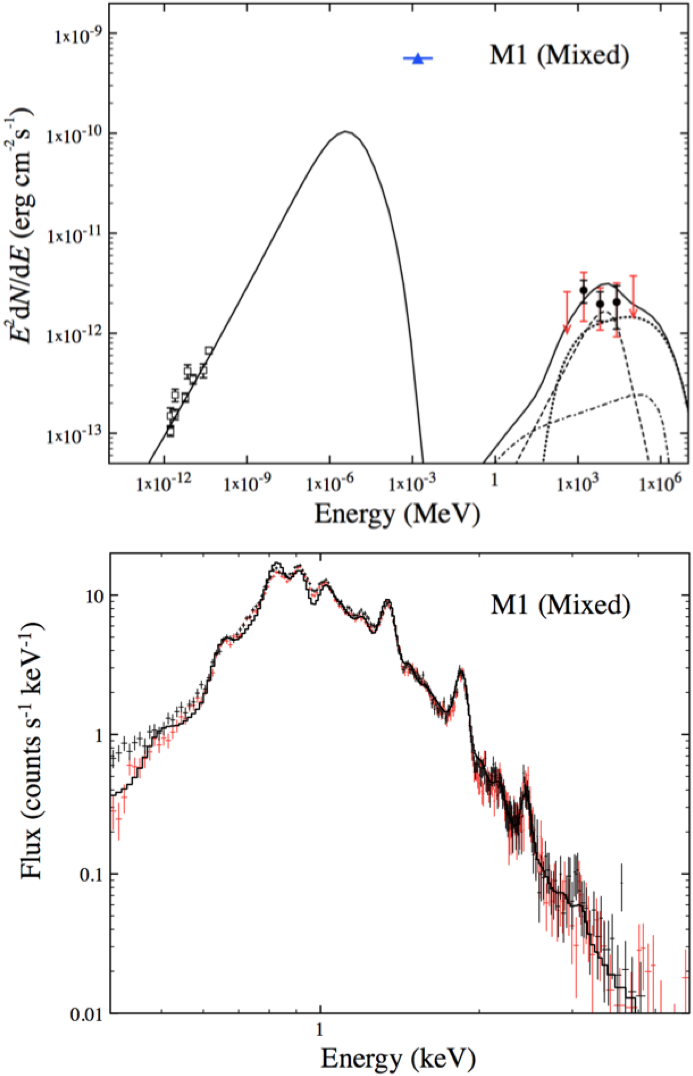}
\caption{ Left: \xmm\ image of CTB~109 (cyan) with \spitzer\ MIPS
24 $\mu$m image of adjacent MC (red).  The white (blue) contours
correspond to the CO line emission as observed by the CGPS, at
velocity -51 (-54)~km~s$^{-1}$, and the levels correspond to 0.5
and 1.5 K. The 95\% confidence radius of the centroid of the
associated Fermi-LAT source is shown as a dashed red circle.  Right:
Broadband emission model for CTB~109 (top) and thermal X-ray emission
fit to predictions of broadband model (bottom). [Figures from Castro
et al. 2012. Used by permission of the AAS.] } \label{ctb109}
\end{figure}

A much more dramatic example of an SNR/MC interaction is that of
CTB~109. As shown in Figure \ref{ctb109} (left), the X-ray emission
(cyan) is characterized by a half-shell morphology, with the western
half of the SNR completely missing. CO measurements reveal a massive
MC in this western region (white contours) from which IR emission
is also seen (red image, from Spitzer). Radio observations at 1420
MHz show the same half-shell morphology \citep{kothes_etal_02},
confirming that the missing X-ray emission in the west is not
produced by excess absorption. Rather, upon encountering the massive
cloud, the SNR shock has apparently completely stalled.

\begin{figure}[t]
\centerline{\includegraphics[height=0.4\textheight]{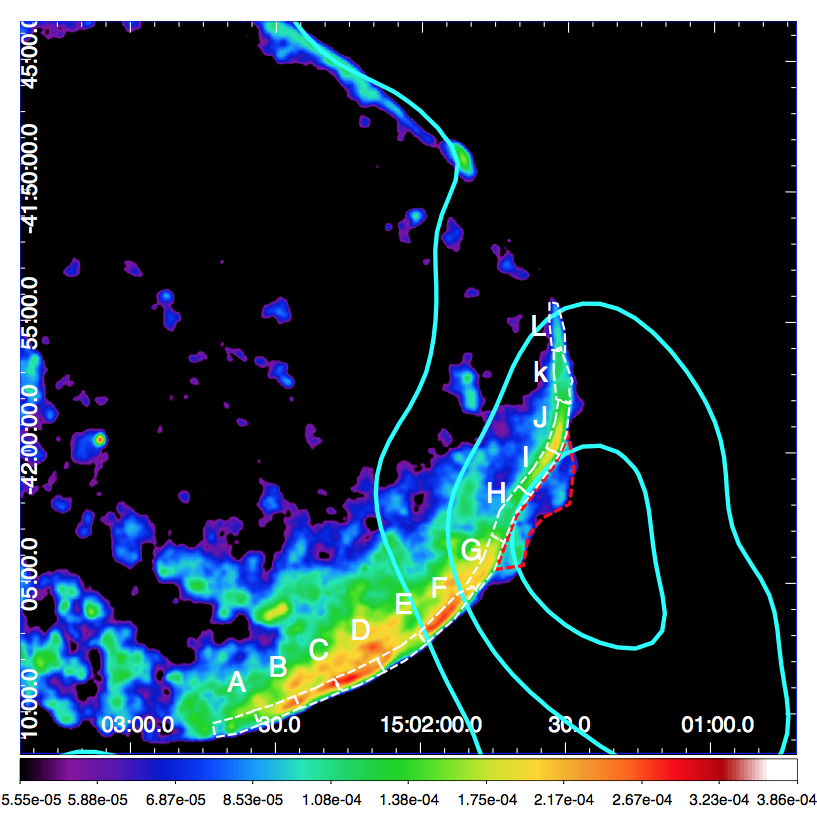}}
\caption{{\it 
XMM-Newton} image of the southwestern portion of SN 1006 in the
0.3-2 keV band with HI column density plotted in contours showing
the presence of a neutral gas cloud perfectly fitting the concavity
in the SNR limb (Figure from Miceli et al. 2014).
}
\label{hi-sn1006}
\end{figure}

A different approach was very recently adopted by \citet{miceli_etal_14},
where the authors demonstrated the connection between shock-cloud
interactions and particle acceleration in the southwestern limb of
the historic remnant of SN 1006 based on the comparison of X-ray
data with HI data. Figure \ref{hi-sn1006} shows the X-ray image of
the SW part of SN1006 in the 0.3-2 keV band with HI contours
superimposed. Exactly at the position where an indentation is
observed in the X-ray (and radio) limb, there is an HI cloud. The
existence of enhanced density was also confirmed by other means by
\citep{winkler_etal_14}.  Several lines of evidence indicate that
at this site  particle acceleration is highly efficient and there
is shock/cloud interaction,  thus making this site a very promising
source of \gamray\ hadronic emission likely to be detectable with
the {\it Fermi} LAT in the near future.

\subsubsection{Mixed Morphology}

\begin{figure}
\includegraphics[height=0.3\textheight]{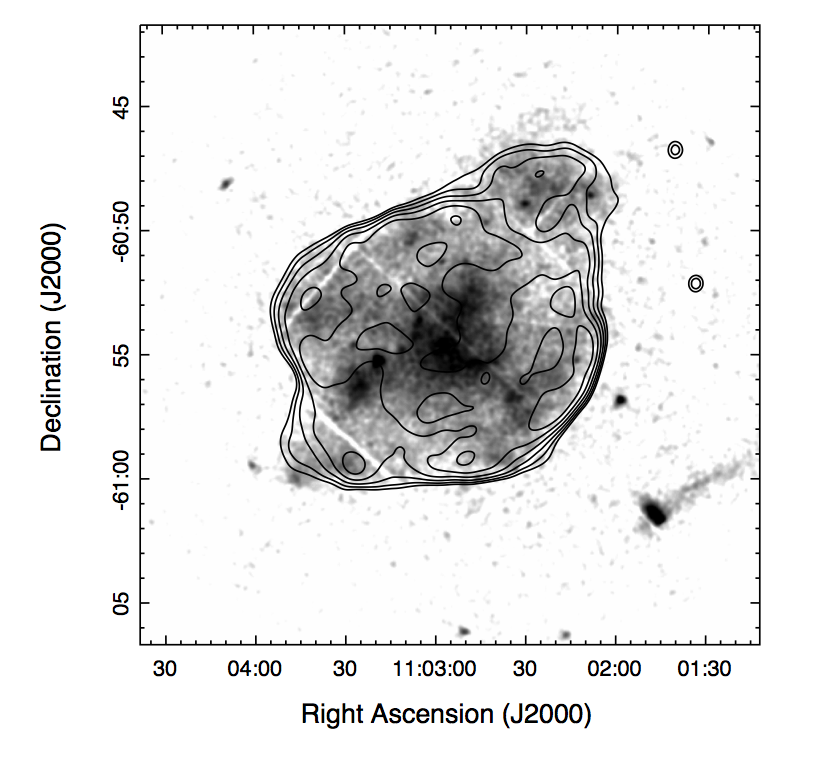}
\includegraphics[height=0.3\textheight]{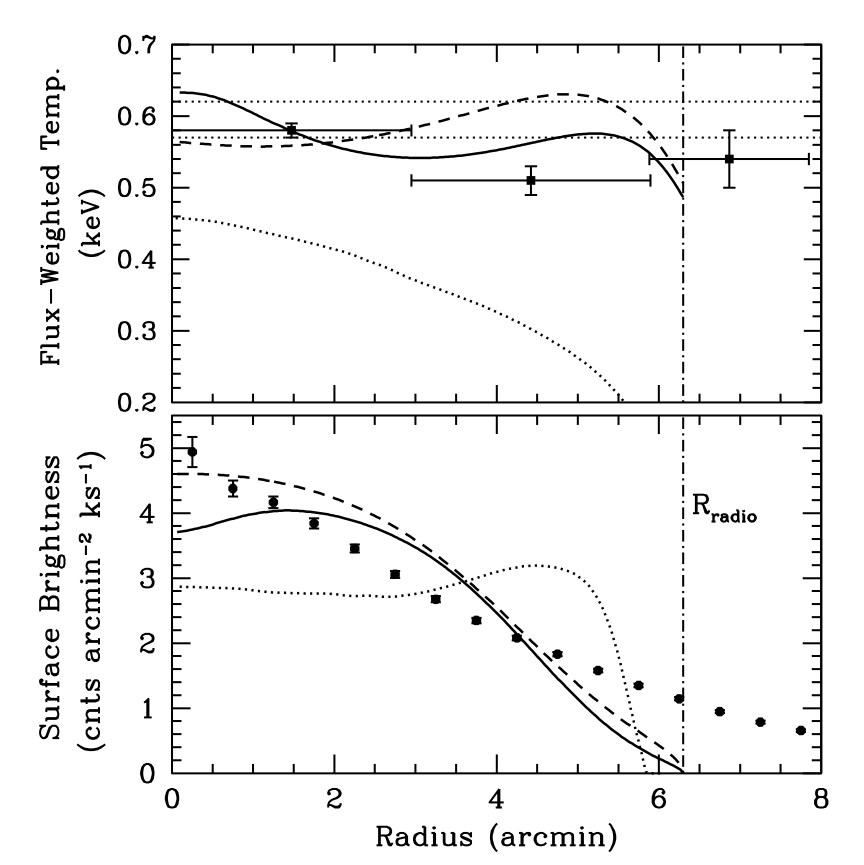}
\caption{
Left: \xmm\ image of MSH~11$-$61{\it A} with radio contours from 
MOST. The centrally-dominant emission is thermal in nature, and
characterizes this as a MM SNR.
Right: Brightness (bottom) and temperature (top) distributions for 
MSH~11$-$61{\it A} along with predictions from cloudy ISM (solid, dashed) 
and thermal conduction (dotted) models (after Slane et al. 2002). 
Horizontal lines in upper panel indicate mean temperature range.
}
\label{mm}
\end{figure}

A distinct subclass of remnants, known as mixed morphology (MM)
SNRs, is characterized by a typical shell-like radio morphology
contrasted by a centrally-bright X-ray morphology in which the
central emission is thermal in nature (Figure \ref{mm}, left). At
present, there are $\sim 40$ known MM SNRs \citep[see summary
by][]{vink12}, but the nature of the central X-ray emission is
poorly understood. While abundance determinations from X-ray spectra
indicate evidence for the presence of ejecta in some such remnants
\citep[e.g.,][]{shelton_etal_04,ls06,bocchino_etal_09,pannuti_etal_10,yamaguchi_etal_12},
the estimated mass of X-ray emitting material in the central regions
is generally much too high to be composed primarily of ejecta.

The spectral properties in MM SNRs provide significant constraints
on the evolution of these systems. The plasma appears to be nearly
isothermal in many systems, quite in contrast to the temperature
profile expected in the Sedov phase of evolution. In addition,
recent studies have shown that the plasma is overionized in several
MM SNRs (see Section 3.2.2). This may result from early evolution
through a dense wind profile created during the late phase of a red
supergiant (or perhaps Wolf-Rayet) progenitor. In this scenario,
the plasma is quickly ionized as the shock passes through the dense
wind, but rapidly cools when the shock breaks through into the lower
density surroundings, leaving the plasma in a higher ionization
state than expected for its temperature \citep{moriya12}. It is not
clear exactly how such a scenario also leads to a centrally-bright
morphology at later stages, nor have detailed models for such a
scenario been carried out to confirm that a sufficient amount of
material is shocked at this early stage to persist as an observable
overionized feature in late stages of evolution.

Early models for MM SNRs centered primarily on two scenarios.  In
one, evolution to the radiative phase in which the
shell temperature drops to low temperatures allows
interstellar absorption to reduce the observed emission from the
shell, while thermal conduction while the remnant is young and hot
results in the transport of heat and bulk material into the central
regions, smoothing out the temperature profile and increasing the
emission from the remnant center \citep{cox_etal_99}. Application of
this model to W44 is able to reproduce some general features, but
fails to fully explain both the temperature and brightness distributions
observed for this remnant \citep{shelton_etal_99}.

A second scenario centers on evolution in a surrounding medium
filled with dense clouds \citep{white_long_91}. Upon being overtaken
by the expanding SNR, the clouds evaporate through saturated
conduction in the hot remnant interior, slowly increasing the central
emission.  For different combinations of the cloud evaporation
timescale and the ratio of the mass in clouds to that in the
intercloud material, significantly peaked brightness profiles can
be obtained.  Applying this model to the emission from MSH~11$-$61{\it
A}, \citet{slane_etal_02} found that reasonable agreement could
be obtained for both the radial brightness and temperature distributions
(Figure \ref{mm}, right), but that the required evaporation timescale
($\sim 10 - 40$ times the age of the remnant) appears much longer
than expected.  Application of the thermal-conduction/radiative-phase
scenario was unable to reproduce the temperature and brightness
distributions.

Importantly, the vast majority of MM SNRs appear to be interacting
with MCs, and nearly half are observed to produce \gamray\ emission,
providing an additional clue as to the conditions that lead to the
observed properties.  At this stage, though, MM SNRs remain poorly
understood. At the same time, increases in our knowledge of the
abundance, ionization, and thermal properties of the shocked
central plasma offer promise for constraining more complete models
of these systems.

\subsection{Spectral Signatures}

The temperature, composition, and ionization state of the shocked gas 
in an SNR all depend crucially on the properties of the medium into
which the remnant evolves. In particular, dense environments associated
with the presence of MCs can produce spectral signatures that reveal
the SNR/MC interactions. Such signatures are readily observed in
a host of individual remnants.

\subsubsection{Temperature and Absorption Variations} 

As an SNR sweeps up excessive amounts of material in the dense
regions around a MC, the shock velocity drops. If the MC interaction
is confined to only discrete regions in the SNR, this can result
in temperature variations in the shocked plasma.  X-ray studies of
W51C \citep{koo_etal_2002,koo_etal_2005} reveal a $\sim 20\%$
temperature decrease in the northern regions where the SNR is
interacting with a MC. Spectral modeling also indicates an increase
in the column density, $N_H$, in this region, suggesting that a
portion of the MC lies between the remnant and observer.

X-ray studies of 3C~391 also reveal $N_H$ variations associated
with a MC interaction, here in the northwestern regions of the
remnant \citep{cs01}. The increase of $\sim 5 \times 10^{21}{\rm\
cm}^{-2}$ indicates a molecular cloud density of $\langle N(H_2) \rangle \sim 100
l_{pc} {\rm\ cm}^{-3}$ where $l_{pc}$ is the depth, in pc, of the
MC region residing in front of the SNR.

\subsubsection{Ionization Signatures}

Due to the high densities for SNRs encountering MCs, one expects
the shocked plasma to quickly reach ionization equilibrium.  This
is indeed observed in some remnants. For example, along the northeast
limb of W28, \citet{nakamura_etal_14} find that the plasma is in
collisional ionization equilibrium (CIE), with signs of density
variations from region to region. In the central regions of the
remnant, radiative recombination continuum (RRC) features of He-like
Si and S are observed \citep{sk12}, indicating a recombining plasma.
Further, excess emission near the K$\alpha$ lines of He-like Mg
and Ne indicate different ionization temperatures for different
elements.

Such overionization states are also observed in several other SNRs
for which MC interactions occur. IC~443 shows an enhanced intensity
ratio for H-like to He-like Si, indicating an overionized plasma
\citep{kawasaki_etal_02}, and also distinct RRC features
\citep{yamaguchi_etal_09}, for example. RRC features are also
observed for W44, which also shows enhanced emission from H-like
Si \citep{uchida_etal_12}, and from W49B \citep{ozawa_etal_09}.
Interestingly, the nature of such overionized plasmas may actually
be only indirectly associated with the presence of MCs. Although
thermal conduction has been suggested as a mechanism by which heat
flow from the hot SNR interior to regions with cold clouds could
bring lower the electron temperature to values below the current
ionization temperature of the ions, most calculations indicate that
this process is too slow to operate efficiently in the typical
lifetime of an SNR \citep[e.g.,][]{uchida_etal_12}. Instead, early
evolution through a dense medium such as that from a stellar wind
may have created a high ionization state, with subsequent rapid
cooling as the remnant expands adiabatically into a low density
cavity leading to a temperature that is lower than the ionization
temperature \citep[e.g.,][]{yamaguchi_etal_09,uchida_etal_12,moriya12}.
Maps of the ionization state in W49B indicate higher states of
overionization in the central and western regions
\citep{miceli2010,lopez_etal_13}, supporting the notion that rapid
expansion in the direction opposite that of a dense MC in the east
is responsible for the cooling.

Given the above scenario, it would seem that the most direct
connection between recombining plasma in SNRs and their association
with MCs is simply that the proposed early-phase evolution through
a dense stellar wind implies a massive progenitor star, and the
remnants of such stars are often found near the molecular cloud
complexes from which they formed. However, it also appears that the
overionized SNRs are all of the mixed morphology class. Whether
or not the overall temperature, ionization, and brightness properties
of this class require specific contributions from MC interactions
remains unknown, at present, but is an area of active study.

\subsection{Abundances and Nonthermal Emission}

In evolved remnants that have undergone considerable interactions
with MCs, the total swept-up mass can be very large. If the shock
velocity in the dense interaction regions is low, much of this
material will be too cool to produce X-rays. However, in many systems
the X-ray emitting mass, $M_x$, is also large. Modeling of the
emission from W44 indicates roughly $100 M_\odot$ of hot gas in the
remnant interior \citep{shelton_etal_04}, for example. The roughly solar
abundances for this swept-up material will thus act to severely
dilute the enhanced abundances of any (much smaller) ejecta
component. Nonetheless, traces of ejecta appear common in many mixed
morphology remnants \citep[e.g.,][]{slane_etal_02,shelton_etal_04,ls06,
bocchino_etal_09}, virtually all of which appear to be interacting
with molecular clouds. In IC~443, a distinct ring-like structure of
hot ($\sim 1.4$~keV) plasma with significantly enhanced abundances
of Mg and Si is observed in the vicinity of a MC interaction region,
suggesting that a strong reverse shock produced in this interaction
with dense material has produced enhanced ejecta emission 
\citep{troja_etal_08}. 

Evidence for ejecta in IC~443 also exists in the form of compact
knots of X-ray emitting material \citep{bb03}. The observed hard
spectra and enhanced abundances suggest that these may be fast-moving
knots of ejecta traveling into dense molecular material and producing
$K\alpha$ emission accompanied by nonthermal bremsstrahlung (NTB)
emission from shock-accelerated electrons \citep{bykov02}. Knots
with similar spectral properties are observed in Kes~69 and are
coincident with CO emission from a nearby MC \citep{bocchino_etal_12},
reinforcing this interpretation. Discrete X-ray knots directly along
the SNR/MC interaction region in 3C~391 are also observed \citep{chen_etal_04},
but while the inferred density for these knots is high, consistent
with structures being driven into clouds, the abundances do not
show strong evidence for metal enhancements. We note that a
complete analysis of the composition of cold, dense, metal-rich
ejecta from X-ray observations needs to also account for the
resonant photo-absorption effect.

Regions of {\it low} abundance plasma have also been observed in
some regions of SNRs interacting with MCs.  The Fe abundance appears
to be subsolar in some bright knots in W44. The abundances of other
elements in these knots appear to be enhanced, possibly indicating
a scenario in which ejecta fragments are traveling through MC
material in which some Fe remains condensed onto dust grains
\citep{shelton_etal_04}.  In W28, the emission from the bright
northeast rim that is adjacent to a MC is best described by a high
density plasma in CIE with subsolar abundances of refractory elements
\citep{nakamura_etal_14}, but in this case the abundances of volatile
elements are also subsolar, complicating any interpretation of
condensation onto dust grains in MC material. Similar depletions
of N, O, and Ne are observed in the XA region of the Cygnus Loop
\citep{mb11}.

\section{Gamma-rays from SNR/MC Interactions}

\begin{figure}[t]
\centerline{\includegraphics[width=0.7\textwidth]{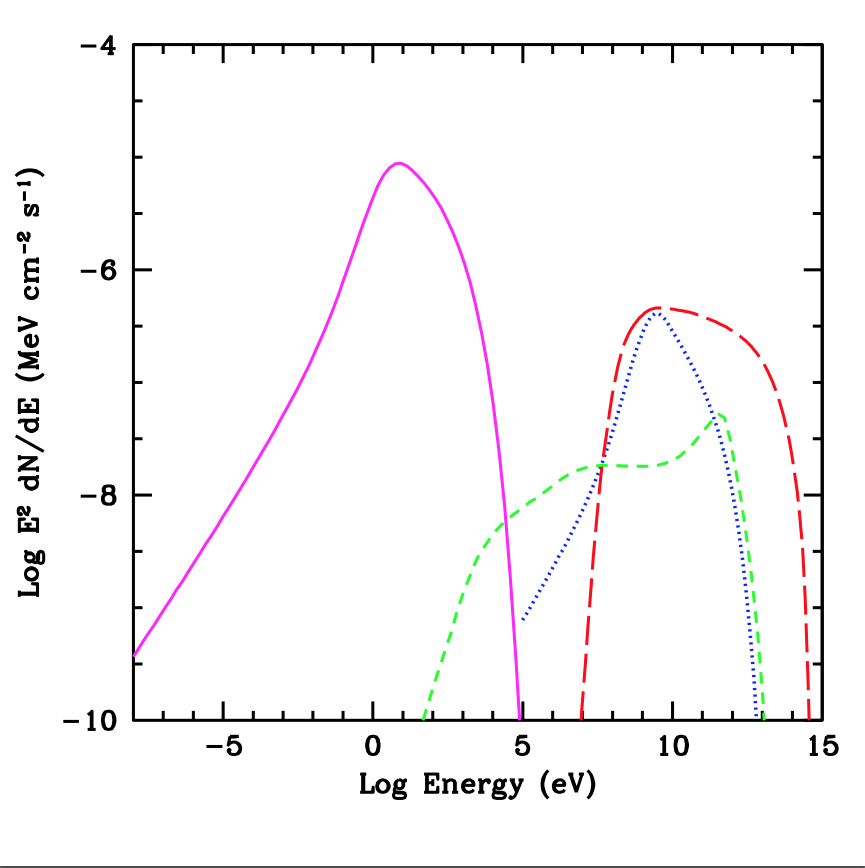}}
\caption{
Simulated broadband spectrum from SNR undergoing efficient DSA
of electrons and protons. The solid magenta curve represents
synchrotron emission, the dotted blue curve is IC emission, the
dashed green curve corresponds to NTB, and the long-dashed curve
represents the $\pi^0$-day emission.
}
\label{gray}
\end{figure}

\begin{figure}[t]
\centerline{\includegraphics[width=1.0\textwidth]{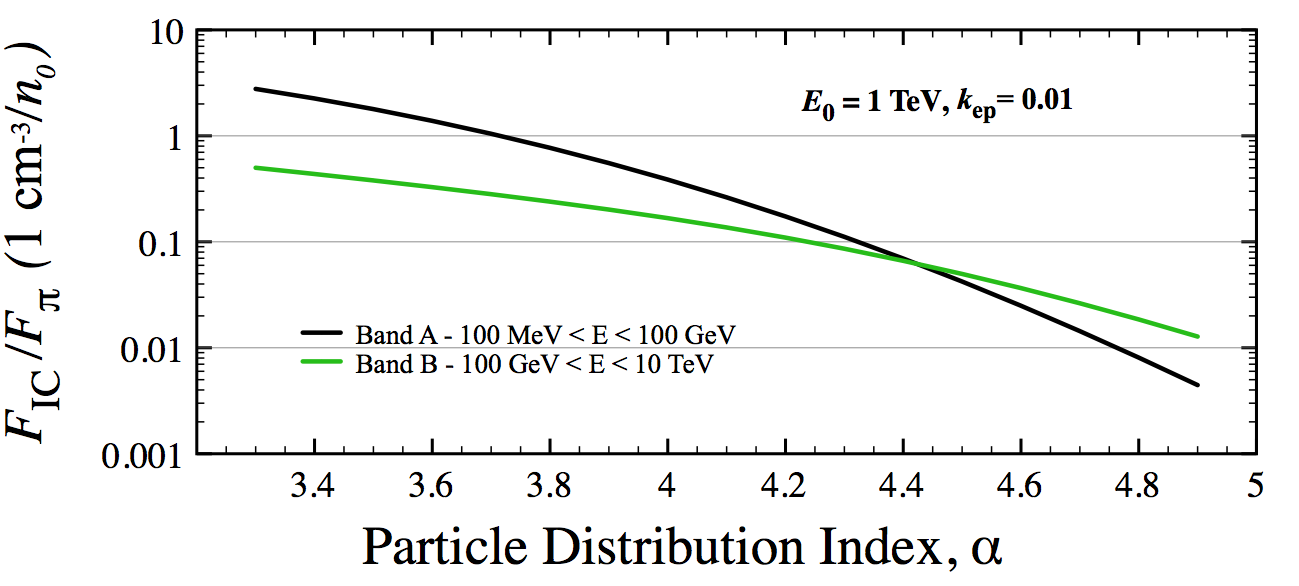}}
\centerline{\includegraphics[width=1.0\textwidth]{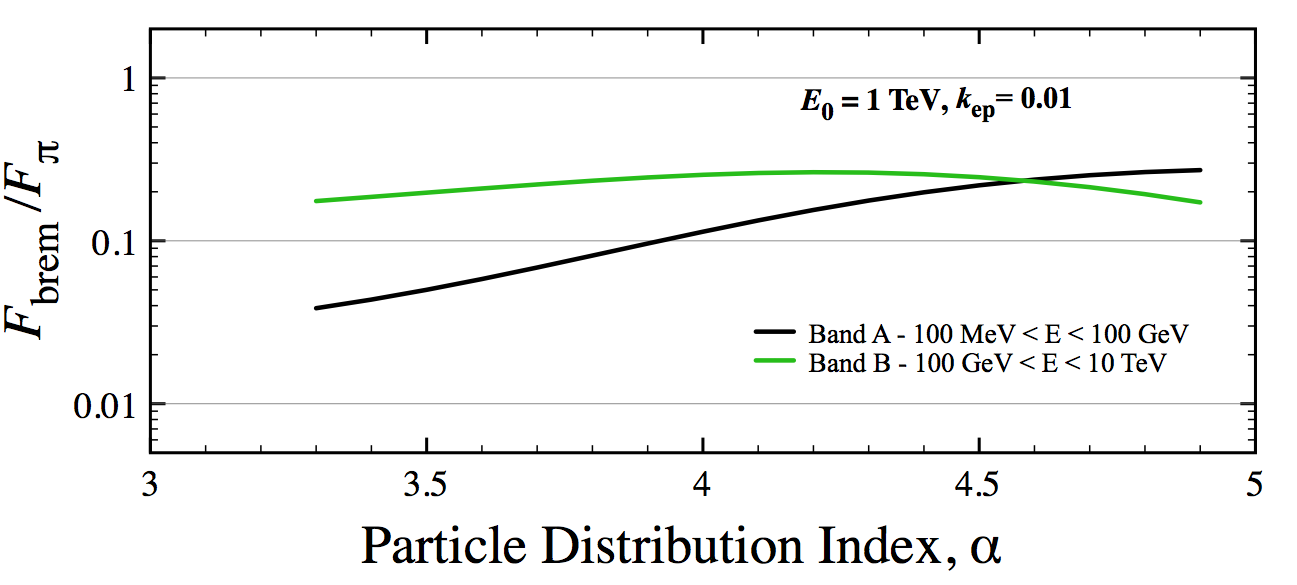}}
\caption{
IC-to-$\pi^0$-decay (top) and
NTB-to-$\pi^0$-decay (bottom) flux ratios as a
function of the particle momentum distribution spectral index,
$\alpha$. The black curve is for the photon energy band 100 MeV to
100 GeV, and the green line indicates the flux ratio in the 100 GeV
to 10 TeV range. The exponential particle energy cutoff, for both
electrons and protons, has been fixed at 1 TeV, and the electron-to-proton
ratio is $K_{ep} = 0.01$. [From Castro et al. 2013.  Reproduced by
permission of the AAS.]
}
\label{pi0}
\end{figure}

\subsection{Particle Acceleration in SNRs}

\noindent Particle acceleration in SNRs has long been suggested as
a major contributor to the cosmic ray population, at least up to
the knee in the spectrum at $\sim 10^{15}$~eV. The generally assumed
mechanism is DSA, where some particles scatter off of self-generated
turbulence and are returned to the shock region multiple times.  In
nonlinear DSA, a non-negligible fraction of the electrons and ions
at the forward shock (FS) can reach ultrarelativistic energies.
Diffusive shock acceleration has received considerable attention,
and specific predictions of the nonlinear theory include
\citep[e.g.,][]{BE87,MD2001,bbmo13}: (i) accelerated particles
obtain enough pressure to modify the shock structure, with the shock
developing an extended upstream precursor; (ii) the overall shock
compression ratio can increase above the Rankine-Hugoniot value of
four for strong shocks, while simultaneously the subshock compression,
which is mainly responsible for heating the unshocked plasma,
decreases below the Rankine-Hugoniot value; (iii) the particle power
law expected from test-particle DSA becomes concave, with the highest
energy particles developing a spectrum harder than the test-particle
power law; (iv) a significant fraction of the highest energy particles
can escape from the shock, adding to the nonlinear nature of the
mechanism; and (v) the production of superthermal particles goes
hand-in-hand with the production of magnetic turbulence, and strong
magnetic field amplification can occur in high Mach number shocks.

While there is no direct evidence for the specific mechanism of DSA
in the forward and reverse shocks of SNRs, there is overwhelming
evidence for the production of relativistic particles, either ions
or electrons, at the forward shocks in a number of remnants. There
is also clear evidence for magnetic field amplification, and for
the modification of the plasma hydrodynamics, predicted by efficient
DSA, in several remnants
\citep[e.g.,][]{pf07,uchiyama_etal_07,Cassam2008,ua08} This, combined
with the direct evidence for efficient DSA in spacecraft observations
of heliospheric shocks \citep[e.g.,][]{EMP90} and confirmation of
fundamental aspects of the theory from particle-in-cell (PIC)
simulations \citep[e.g.,][]{EGBS93,SSA2013}, has added to a general
acceptance of the mechanism for cosmic ray production in SNRs.

Despite the progress made in understanding DSA and verifying some
of its basic predictions, important open questions remain about the
maximum particle energy \Emax, the acceleration efficiency, the
electron-to-proton ratio $\Kep$ for the injected particles, and the
eventual escape of cosmic rays from the acceleration region.  As
an example, local cosmic ray measurements, as well as theoretical
expectations, suggest that shocks put far more energy in protons
and heavier ions than in electrons. Yet the nonthermal emission
from most astrophysical sources is dominated by radiation from
relativistic electrons. While this can be understood, in part,
because relativistic electrons radiate more efficiently than protons,
the underlying energy budget of a source cannot be determined until
constraints on the hadronic component are obtained and the accelerated
electron-to-proton ratio determined. This is important for cosmic
rays and for all sources where DSA is expected to occur.

Molecular clouds play a particularly important role in this regard
because shocks in the high density MC environment will predominately
produce \gamray s by hadronic interactions rather than leptonic ones
\citep[e.g.,][]{adv94}.

\subsection{Gamma-ray Production}

Gamma-ray production from such relativistic particles can proceed
through inverse-Compton (IC) scattering of ambient photons by the
energetic electrons, nonthermal bremsstrahlung (NTB) from collisions
between relativistic electrons and ambient material, and the decay
of neutral pions formed in collisions of energetic protons with
surrounding nuclei.  Figure \ref{gray} presents a simulation of the
broadband spectrum produced by an SNR undergoing efficient acceleration
of electrons and ions. The magenta curve represents synchrotron
emission from the relativistic electrons, and the dotted blue curve
corresponds to IC emission produced from that same electron population
up-scattering photons from the cosmic microwave background (CMB).
The dashed green curve represents NTB from the relativistic electrons
interacting with ambient material, and the long-dashed red curve
corresponds to \gamray s from the decay of neutral pions produced
by collisions of the relativistic proton component with ambient
nuclei.  (The curvature of the particle spectra associated with DSA
is particularly evident in the synchrotron and IC emission.)

The emission from both NTB and $\pi^0$-decay scales with the ambient
density $n_0$. As a result, in high density environments such as
those encountered in SNR/MC interactions, significant \gamray\
emission is expected if the SNR has been an active particle
accelerator. For $\Kep \sim 10^{-2}$, as measured locally, the
$\pi^0$-decay emission will dominate (see Figure \ref{pi0}), making
such \gamray\ emission an important probe of the hadronic component
of the particles accelerated by the SNR.

An important additional consideration for \gamray\ production is
the local photon energy density. Contributions from starlight as
well as IR emission from local dust can increase the IC emission
and change the spectral shape of this component due to the different
effective temperatures of these photon components.  These contributions
are highly dependent upon galactocentric radius
\citep[e.g.,][]{strong_etal_2000}. Moreover, IR emission produced
by the SNRs themselves can contribute significantly to the IC
\gamray\ emission \citep[e.g.,][]{MC2012,slane_etal_2014}.

Modeling of the broadband spectra from such SNRs, in order to
ascertain the nature of the $\gamma$-ray emission, is complicated
and has led to mixed interpretations, making the evidence for ion
acceleration controversial in some cases. However, in a growing
number of cases, $\gamma$-ray emission from some SNRs known to be
interacting with MCs seems to clearly require a significant component
from pion decay. We discuss emission from several of these sources
below.

\subsection{Gamma-ray Observations of SNRs} 
\noindent 
To date, studies have identified \gamray\ emission from $\sim 25$
SNRs, the majority of which are interacting with MCs.  Of these,
the evidence for energetic hadrons as the source of these \gamray
s is compelling for more than 50\% of these based on energetic
arguments and/or broadband spectral modeling.  \fermi\ LAT observations
of W51C reveal a spectrum that is consistent with $\pi^0$-decay,
with dominant NTB ruled out unless $\Kep \gg 0.01$, and IC emission
ruled out on energetic grounds \citep{AbdoEtalW51C2009}.  W44 and
IC~443 show clear evidence of a kinematic ``pion bump'' in their
spectra, firmly establishing the presence of energetic ions in these
remnants \citep{AbdoEtalW442010,GiulianiEtal2011,Ackermann_Science}.
Gamma-ray emission from CTB~109 (Figure \ref{ctb109}, left) has
been detected with the \fermi\ LAT \citep{castro_etal_12}, and
modeling of the broadband emission (Figure \ref{ctb109}, upper
right) indicates that the \gamray\ emission arises from approximately
equal contributions from IC scattering and $\pi^0$-decay, a result
that is strongly constrained by the observed ionization state of
the thermal X-ray emission (Figure \ref{ctb109}, lower right).

Castro \& Slane (2010) carried out {\it Fermi}~LAT studies of a set
of four SNRs (G349.7+0.2, CTB~37A, 3C~391, and G8.7$-$0.1) known
to be interacting with MCs based on observations of hydroxyl (OH)
maser emission at 1720 MHz, and showed that all four were sources
of GeV $\gamma$-ray emission.  Based on the assumption that the
$\gamma$-ray emission is dominated by $\pi^0$-decay produced in the
compressed shell of the SNR, they derived a lower limit on the
density of the $\gamma$-ray emitting material for each remnant, and
compared this with the density inferred from X-ray measurements.
They found that the density inferred from $\gamma$-ray measurements
exceeds that from X-ray measurements by a factor of 20 or more.
Subsequent studies of W41, MSH 17$-$39, and G337.7$-$0.1
\citep{castro_etal_13} as well as Kes~79 \citep{auchettl_etal_14}
reveal \gamray\ spectra indicative of hadronic emission, with
leptonic scenarios requiring total electron energies in excess of
$10^{51}$~erg or $\Kep \gg 0.01$.  Similar discrepancies between
densities inferred from $\gamma$-ray and X-ray measurements are
obtained for these sources.  The magnetic fields implied by the
radio emission in these studies is typically much larger than
expected from the compressed ISM, suggesting evidence for the
magnetic field amplification expected in efficient particle
acceleration in SNR shocks.

\begin{figure}[t]
\centerline{\includegraphics[height=0.5\textheight]{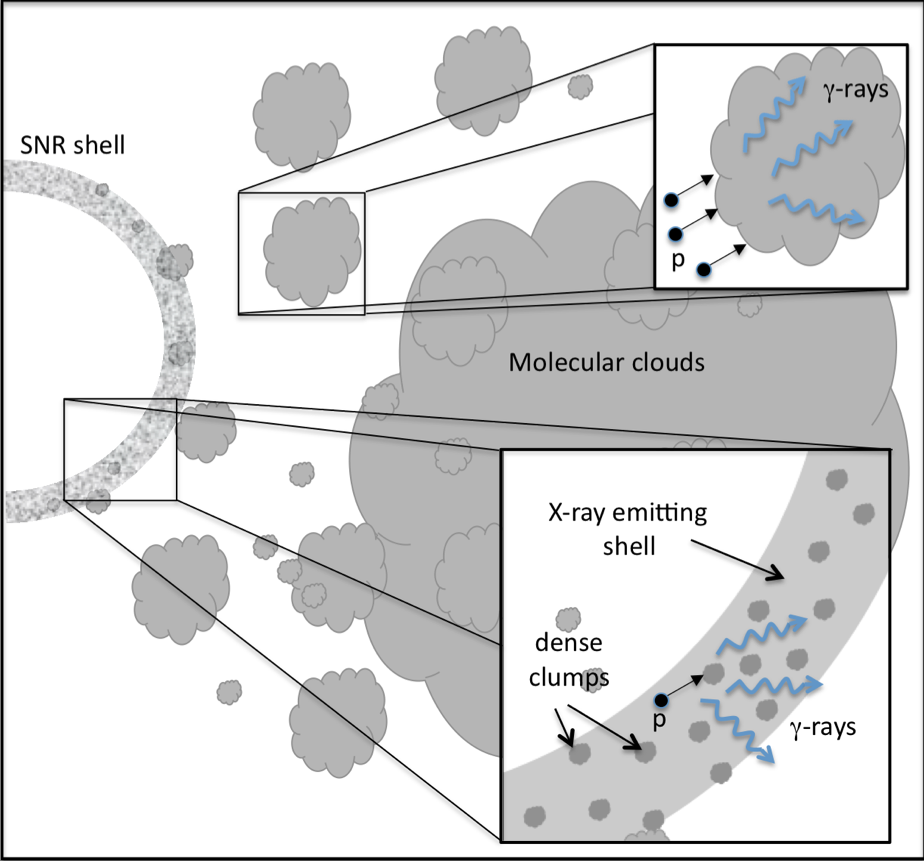}}
\caption{
Schematic diagram of SNR/MC interaction in which postshock region in
SNR contains dense clumps within shocked interclump material. The low-density
shocked interclump gas emits X-rays. Protons acceleration by the SNR
encounter all of gas in the shell, and $\gamma$-rays are produced
primarily through collisions with the dense clumps.
}
\label{cloudy_ism}
\end{figure}

A plausible explanation for the observed discrepancy between $n_x$
and $n_\gamma$  is that these SNRs, by evolving in the presence of
MCs, have swept up clumps of dense material. The SNR blast wave for
each remnant has presumably evolved primarily through the low-density
interclump medium, heating this material to X-ray emitting temperatures
and accelerating particles through DSA. The proton component of
these accelerated particles then interacts with both the dense
clumps and the interclump gas, thus sampling a much higher effective
density than that of the X-ray emitting gas (Figure \ref{cloudy_ism}).
A similar scenario has been proposed by \citep{InoueEtal2012} to
explain the complete lack of thermal X-ray emission from
RX~J1713.7$-$3946, whose observed GeV and TeV $\gamma$-ray emission
is otherwise required to be dominated by IC emission \citep{EPSR2010,
ESPB2012}. Additional modifications of this picture of dense
clumps embedded in the SNR shell invokes reacceleration of cosmic
rays trapped in the cold molecular material rather than 
trapped particles in the shell that have been accelerated by
the SNR \citep[e.g.,][]{uchiyama_etal_10}.

An alternative scenario that can explain \gamray s from SNR/MC
interactions, and also $n_\gamma/n_x \gg 1$, is that of escaping
CRs interacting with external MCs \citep[e.g.,][]{gac09}. Here, as
indicated in Figure \ref{cloudy_ism} (left), there is no connection
between $n_x$ and $n_\gamma$ because the X-rays are produced in the
SNR shell while the \gamray s are from the MC.  Such a scenario has
been suggested to explain the \gamray s from the source MAGIC~J0616+225
as the result of energetic particles that have escaped from IC~443
and are interacting with dense clouds \citep{torres_etal_08}, for
example, and discrete TeV sources outside the remnant W28 have been
suggested to originate from particles escaping the SNR and interacting
with adjacent clouds \citep{aharonian_etal_08}.  The expected
\gamray\ spectrum in this scenario is complicated, however. Because
the escaping particles are distributed around \Emax, the resulting
$\pi^0$-decay spectrum is peaked at higher energies than that for
the shell \citep{zp08,eb11}. However, because \Emax is expected to
decrease with time, the time-accumulated spectrum of particles
interacting with a remote cloud will depend on the age of the SNR
and the energy-dependent diffusion process by which the particles
are transported. For simple assumptions about the time evolution
and particle diffusion, this could lead to power law \citep{gac09}
or broken power law \citep{omy11} particle distributions at the
remote cloud site. The observed \gamray\ spectra from W51C, W28,
W44 and IC 443 show distinctly different slopes, leading to questions
as to whether these differences are intrinsic to the acceleration
or the result of complex evolution, propagation, and interaction
with remote clouds.

\section{Conclusions}
Observations of SNR/MC interactions provide unique information on
the distances to SNRs and on the physics of shock/cloud interactions.
X-ray observations, in particular, probe the state of the shocked
gas and provide diagnostics of the interaction through absorption,
ionization, temperature, and general morphology measurements. For
SNRs that are, or have been, active particle accelerators, these
interactions with dense cloud material can lead to crucial measurements
of emission from the hadronic component of their relativistic
particle spectra, providing crucial information on acceleration
efficiency, maximum energies, and particle escape in these systems.

Recent studies of \gamray\ emission from numerous SNRs demonstrate
that those interacting with MCs show strong evidence of relativistic
protons. For some, X-ray measurements of the flux or ionization
state of the thermal plasma provide constraints that allow us to
determine what fraction of the \gamray\ emission arises from this
hadronic component. For a significant number of the remnants
interacting with MCs, the inferred density of the \gamray -emitting
material is considerably higher than that determined from X-ray
measurements, suggesting a complex emission environment with dense
clumps embedded in hot interclump gas, or perhaps significant
\gamray\ emission associated with escaping cosmic rays interacting
with external clouds. Considerable additional observations and
modeling efforts are required to better understand the the postshock
regions of SNR/MC interactions, as well as the evolution and transport
of the escaping particles.  Of particular importance are simulations
of SNR evolution into dense, cloudy environments to study the
formation of clumps in the postshock region, as well as modeling
of realistic particle acceleration with escape and energy-dependent
diffusion to predict the spectrum of SNR cosmic rays interacting
with nearby MCs.  Equally important is progress in observations and
modeling of supernova ejecta interactions with a complex circumstellar
medium formed by the winds of massive stars at the different evolution
stages of the massive stars.  On the observational side, more
sensitive X-ray and $\gamma$-ray observations are needed to provide
improved spectra of SNRs interacting with MCs.  Improved angular
resolution at $\gamma$-ray energies, such as may be expected in
future facilities like CTA, is particularly important.  In combination
with current studies of plasma instabilities, magnetic field
amplification, and the reacceleration of relic cosmic cosmic rays,
the outlook for reaching a quantitative understanding of the role
SNRs play in producing Galactic cosmic rays is extremely positive.

\noindent

\begin{acknowledgements}
The authors would like to thank Tom Dame and Herman Lee for important
discussions contributing to this paper.  We also thank ISSI and
its staff for supporting this effort and for creating a stimulating
environment for the meeting at which this work was initiated.  POS
acknowledges support from NASA Contract NAS8-03060.  AMB was supported
in part by RAS Presidium 21 and OFN 15 programs.
\end{acknowledgements}

\bibliographystyle{aps-nameyear}       

\bibliography{bib_slane}   

\nocite{*}


%
%

\end{document}